\documentclass[twocolumn,showpacs,superscriptaddress,preprintnumbers,amsmath,amssymbl]{revtex4-2}
\usepackage{amssymb}
\usepackage{graphicx}
\usepackage{dcolumn}
\usepackage{bm}
\usepackage{color}
\usepackage{float}

\begin{document}

\title{Superconductivity in kagome metal ThRu$_3$Si$_2$}
\author{Yi Liu}
\email{liuyiphy@zjut.edu.cn}
\affiliation{School of Physics, Zhejiang University, Hangzhou 310058, China}
\affiliation{Department of Applied Physics, Zhejiang University of Technology, Hangzhou 310023, China}

\author{Jing Li}
\affiliation{School of Physics, Zhejiang University, Hangzhou 310058, China}

\author{Wu-Zhang Yang}
\affiliation{School of Science, Westlake Institute for Advanced Study, Westlake University, Hangzhou 310064, China}%

\author{Jia-Yi Lu}
\affiliation{School of Physics, Zhejiang University, Hangzhou 310058, China}

\author{Bo-Ya Cao}
\affiliation{Department of Applied Physics, Zhejiang University of Technology, Hangzhou 310023, China}

\author{Hua-Xun Li}
\affiliation{School of Physics, Zhejiang University, Hangzhou 310058, China}

\author{Wan-Li Chai}
\affiliation{School of Physics, Zhejiang University, Hangzhou 310058, China}

\author{Si-Qi Wu}
\affiliation{School of Physics, Zhejiang University, Hangzhou 310058, China}

\author{Bai-Zhuo Li}
\affiliation{Department of Physics, Shandong University of Technology, Zibo 255049, China}%

\author{Yun-Lei Sun}
\affiliation{School of Information and Electrical Engineering, Zhejiang University City College, Hangzhou 310015, China}%

\author{Wen-He Jiao}
\affiliation{Department of Applied Physics, Zhejiang University of Technology, Hangzhou 310023, China}%

\author{Wang Cao}
\affiliation{Department of Physics, Shandong University of Technology, Zibo 255049, China}%

\author{Xiao-Feng Xu}
\affiliation{Department of Applied Physics, Zhejiang University of Technology, Hangzhou 310023, China}%

\author{Ren Zhi}
\affiliation{School of Science, Westlake Institute for Advanced Study, Westlake University, Hangzhou 310064, China}%

\author{Guang-Han Cao}
\email{ghcao@zju.edu.cn}
\affiliation{School of Physics, Zhejiang University, Hangzhou 310058, China}
\affiliation{Interdisciplinary Center for Quantum Information, and State Key Laboratory of Silicon and Advanced Semiconductor Materials, Zhejiang University, Hangzhou 310058, China}
\affiliation{Collaborative Innovation Centre of Advanced Microstructures, Nanjing University, Nanjing, 210093, China}

\date{\today}

\begin{abstract}
We report the physical properties of ThRu$_3$Si$_2$ featured with distorted Ru kagome lattice. The combined experiments of resistivity, magnetization and specific heat reveal bulk superconductivity with $T_{\rm{c}}$ = 3.8~K. The specific heat jump and calculated electron-phonon coupling indicate a moderate coupled BCS superconductor. In comparison with LaRu$_3$Si$_2$, the calculated electronic structure in ThRu$_3$Si$_2$ shows an electron-doping effect with electron filling lifted from 100~meV below flat bands to 300~meV above it. This explains the lower superconducting transition temperature and weaker electron correlations observed in ThRu$_3$Si$_2$. Our work suggests the $T_{\rm{c}}$ and electronic correlations in kagome superconductor could have intimate connection with the flat bands.
\end{abstract}
\maketitle

\section{\label{sec:level1}Introduction}
Recently, compounds with kagome lattice have attracted lots of attentions for the unique electronic structure and magnetic frustration~\cite{Review_Ka1,Review_Ka2}, which provides an exciting platform to study rich quantum physics, such as quantum spin liquid~\cite{QSP1,QSP2,QSP3,QSP4}, nontrivial band topology~\cite{TP_Ka1,TP_Ka2,TP_Ka3} and superconductivity~(SC)~\cite{QSP_SC,SC_328,SC_135cal,SC_135,K135,Cs135,GHJ}. One of the most exotic features in the electronic structure of kagome lattice is the flat band with little dispersion in $k$ space, which results from the destructive interference of electron wave functions. The localization of electrons near the Fermi level ($E_{\rm{F}}$) on the flat band are expected to bring high density of electronic states and possible strong electronic correlation~\cite{flat1,flat2}. For superconductivity with conventional phonon-mediated pairing, larger density of states (DOS) usually results in higher critical temperature, while for the case of unconventional non-phonon-mediated paring, the enhanced electron-electron correlation is also prone to yield high-temperature SC. Therefore, it is promising to access flat bands in specialized system to realize higher critical temperature or even unconventional SC.

Recently, new kagome lattice compound CsCr$_3$Sb$_5$, with flat bands closely above the Fermi level, present moderate electronic correlations and unconventional SC under pressure of $\sim$4~GPa~\cite{Cr135}. The mechanism of its unconventional SC and the effect of flat bands are still unclear. Another traditional family of kagome metal with SC and flat bands close to the Fermi level is intermetallic "132"  $RT_3X_2$ phases, where $R$ and $T$ represent a rare-earth element and a transition metal respectively, and $X$ stands for Si or B. $RT_3X_2$ crystalizes in hexagonal layered CeCo$_3$B$_2$ structure, with transition metal $T$ is organized in a distorted kagome layer. Superconductivity is discovered in many materials of $RT_3X_2$ family, such as ThRu$_3$B$_2$, LaRh$_3$B$_2$,  LuOs$_3$B$_2$, LaIr$_3$B$_2$, ThIr$_3$B$_2$~\cite{B132}, YRu$_3$Si$_2$, LaRu$_3$Si$_2$, ThRu$_3$Si$_2$~\cite{Si132}, CeRu$_3$Si$_2$~\cite{Ce132_1,Ce132_2} and LaIr$_3$Ga$_2$~\cite{Ga132}. Recently, LaRu$_3$Si$_2$ with the highest $T_{\rm{c}}$ in $RT_3X_2$ family, attracted much attention for the interesting electronic structure with flat band close to the $E_{\rm{F}}$ and associated strong correlation~\cite{2011La132,2021La132,NQRLa132}. As revealed by the DFT calculation, the flat band is 100~meV above the $E_{\rm{F}}$. Thus, it is possible to tune the flat bands to the Fermi level by introducing extra electrons to achieve a maximum impact on  superconductivity and electron correlations~\cite{2021La132}. The structure and superconductivity of ThRu$_3$Si$_2$ have been reported 30 years ago~\cite{XRD132,Si132}, but the physical properties of normal state and superconductivity have not been systematically investigated yet. Compared with La, Th has similar ion radius, but tends to contribute one more electron. Hence, it's necessary to give a detailed research on this kagome superconductor.

In this work, we carried out a detailed investigation on the superconductivity and normal state of  ThRu$_3$Si$_2$. The measurements of resistivity, magnetism and specific heat demonstrate bulk SC below 3.8~K. Our experimental results suggest a moderate coupled BCS superconductor with electron correlations. Besides, compared with LaRu$_3$Si$_2$, the calculated band structure using first-principles density functional-theory (DFT) shows electron-doping effect, with  the Fermi level shift from below the flat band to above it.
\begin{figure}{H!}
\includegraphics[width=1\columnwidth]{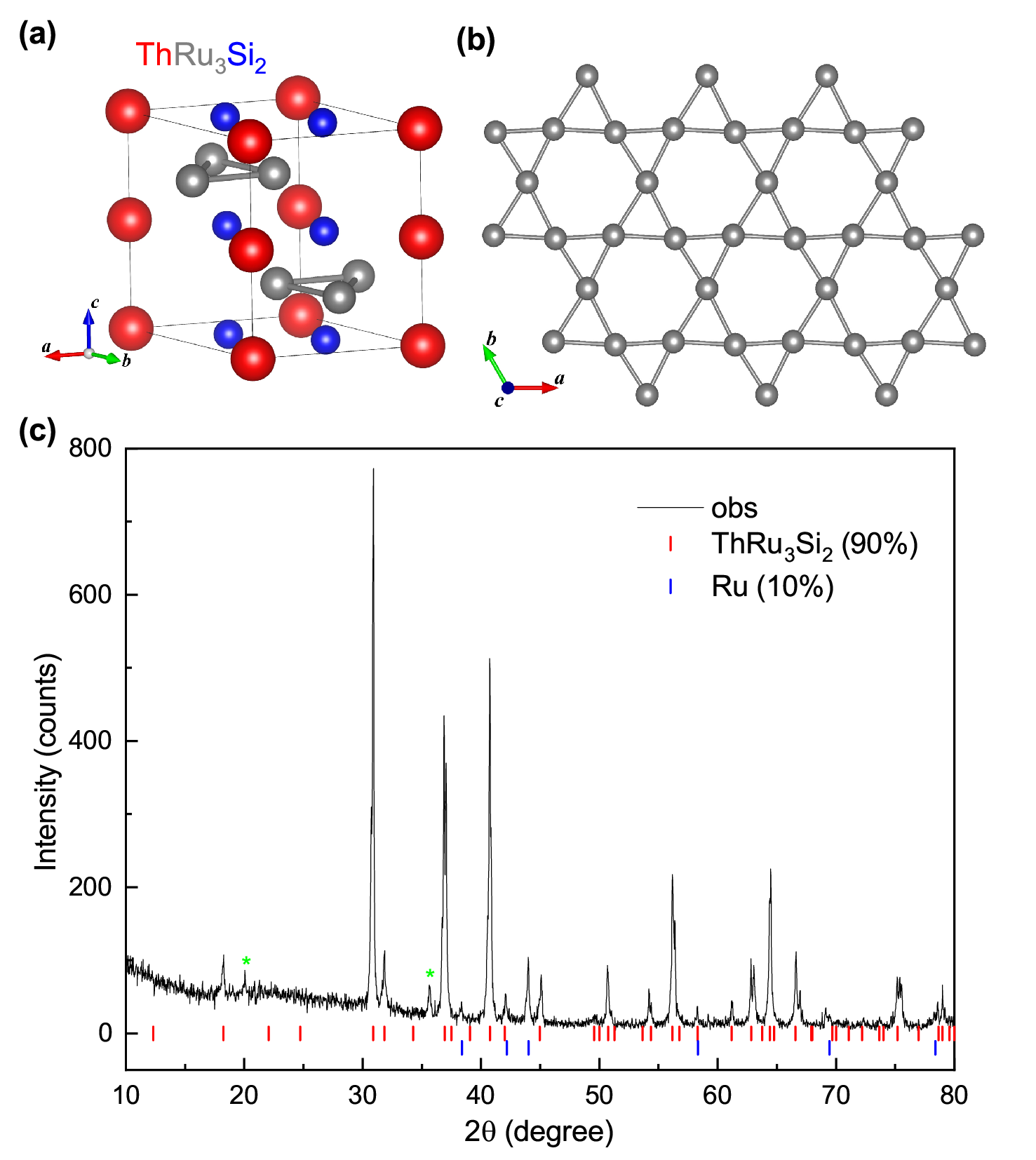}
\caption{(a) Crystal structure of ThRu$_2$Si$_2$, which is formed by the stack of Th(red), Si(blue) and Ru(grey) layers along $c$ axis. (b) Top view of the distorted kagome lattice of Ru atoms. (c) Powder x-ray-diffraction pattern of ThRu$_3$Si$_2$ sample. The molar ratio of ThRu$_3$Si$_2$ and impurity phase of Ru is about 9:1. There are two unidentified small peaks labelled with $\ast$.}
\label{Fig1}
\end{figure}
\section{\label{sec:level2}Experimental Methods}
ThRu$_3$Si$_2$ polycrystal was synthesized through arc melting. Firstly, the high-purity thorium metal was extracted from thorium oxide as previous reported~\cite{Th}. The prepared Th button was scraped into powder and then mixed with Ru~(99.9$\%$) and Si~(99.9$\%$) powder. To avoid the formation of ThRu$_2$Si$_2$, 15$\%$ extra Ru powder is added. The mixture was ground thoroughly and pressed into a pellet, which was then welded in arc furnace with current of 45~A. The process was repeated twice to achieve the uniformity. The compositions of the as-prepared sample is measured by energy-dispersive x ray spectroscopy (EDS) on a scanning electron microscope
(Hitachi S-3700N) equipped with an Oxford Instruments XMax spectrometer, and the determined molar ratio is Th : Ru : Si = 1 : 3 : 2.1. X-ray diffraction (XRD) was carried out on a PANalytical X-ray diffractometer (Model EMPYREAN) with a monochromatic Cu-$K_{\alpha1}$ radiation at room temperature. The diffraction peaks was indexed by ThRu$_3$Si$_2$ and Ru phases, with the percentage of Ru below 10$\%$ concluded from the intensity ratio of the maximum peaks. As shown in Fig.~\ref{Fig1}, the main diffraction peaks can be indexed by the $P6_3/m$ space group (No. 176) with $a$ = 5.613(2) $\rm{\AA}$ and $c$ = 7.190(1) $\rm{\AA}$, which is consistent with the previous report~\cite{XRD132}. Electrical resistivity was measured with four-electrode technique on a brick-shaped polycrystal, while the heat capacity measurement employed thermal relaxation method with a Quantum Design physical property measurement system (PPMS-9). The magnetization measurements were performed on SQUID-VSM magnetometer (MPMS-5).

\section{\label{sec:level3}Results}
\begin{figure*}
\includegraphics[width=1.5\columnwidth]{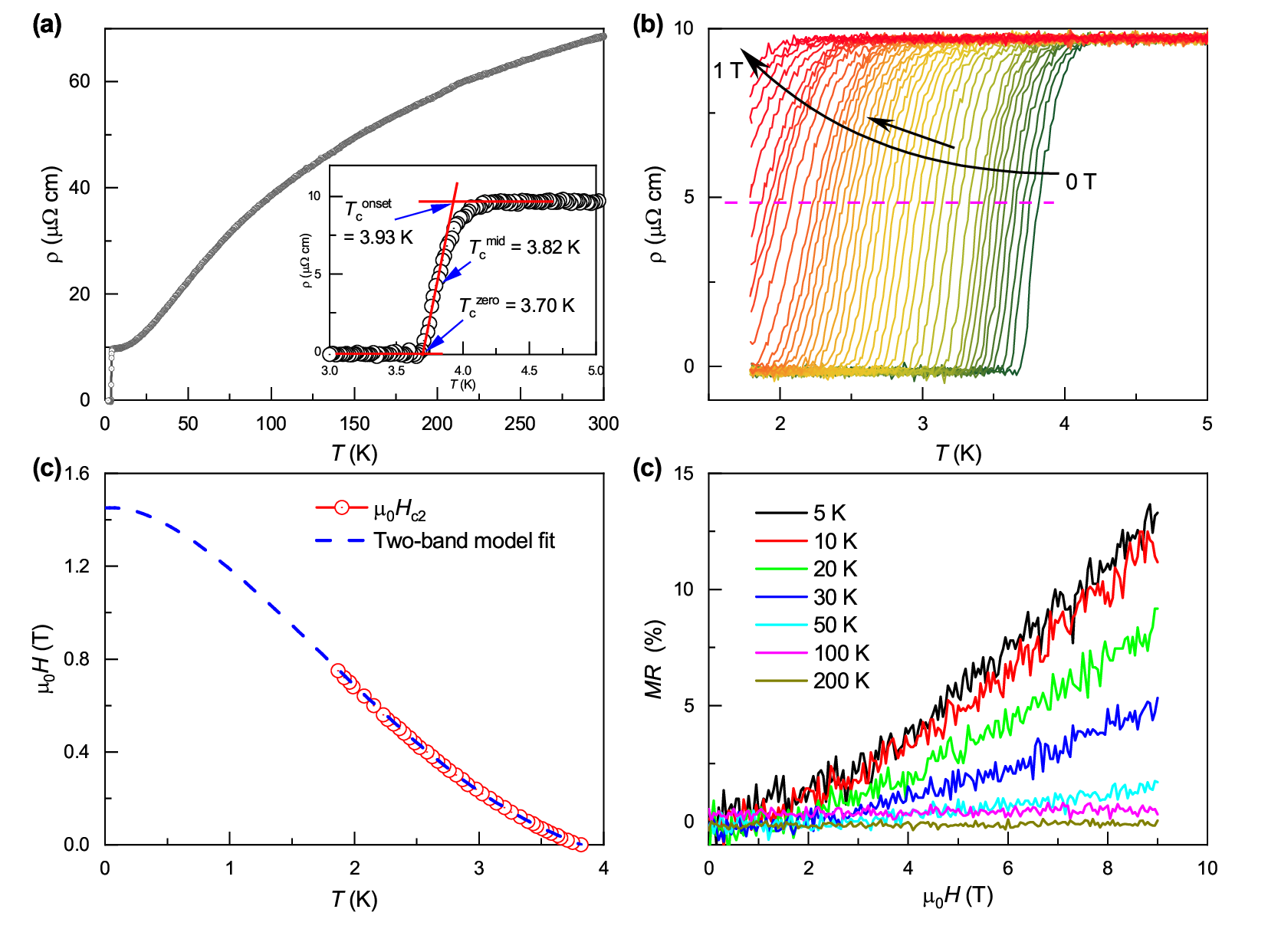}
\caption{(a) Temperature dependence of resistivity for ThRu$_3$Si$_2$. The inset shows the definitions of onset, middle and zero points of the resistivity drop.  (b) Resistivity transition from 1.8~K to 5~K under different magnetic fields from 0 to 1~T. The horizontal dashed pink line locates at half of the normal resistivity to determine the superconducting transition temperatures under fields. (c) Temperature dependence of the upper critical field $\mu_0H_{\rm{c2}}$($T$). The dashed blue line is the fit of two-band model in dirty limit. (d) Magnetoresistance measured at selected temperatures.}
\label{Fig2}
\end{figure*}

The resistivity shows metallic behavior with a large hump below 300~K as present in Fig.~\ref{Fig2}(a), which is similar to the other $RT_3X_2$ compounds~\cite{Ce132_2,2011La132,Y132,Ga132}. The absolute resistivity at room-temperature is about 68 $\mu\Omega$ cm and the residual resistivity ratio RRR = $\rho_{300~{\rm{K}}}$/$\rho_{4~{\rm{K}}}$ reaches 7. Note that, the resistivity and RRR is mainly affected by grain boundary effect in polycrystal instead of its intrinsic conductivity. Superconductivity is observed at low temperature with onset at 3.9~K and zero point at 3.7~K. We define the superconducting transition temperature , $T_{\rm{c}}$ = 3.8~K, which corresponding the middle point where the resistivity is half of that in normal state. Under external fields, the transition is suppressed to lower temperatures and the temperature dependence of the upper-critical field $\mu_0H_{\rm{c2}}$($T$) is shown in Fig.2 (c). The $\mu_0H_{\rm{c2}}$($T$) can be well fitted by two-band model~\cite{two_band}. The fitted  parameters of diffusivity ratio $\eta$, intra-band scattering $\lambda_1$, $\lambda_2$  and inter-band scattering $\lambda_{12}$ = $\lambda_{21}$  of the two bands are 25.27(4), 0.80(1), 0.12(2) and 0.28(1) respectively. The upper-critical field at zero temperature $H_{c2}$(0) is 1.45~T, which is far below the Pauli limit $\mu_0H_{\rm{c2}^{\rm{P}}}$ = 1.84$T_c$ = 6.95~T, suggesting the orbital pairing-broken mechanism is prominent in ThRu$_3$Si$_2$. We also measured the magnetoresistance up to 9~T at selected temperatures, as shown in Fig.~\ref{Fig2}(d). The magnetoresistance is about 13$\%$ at 5~K and 9~T.

\begin{figure*}
\includegraphics[width=1.5\columnwidth]{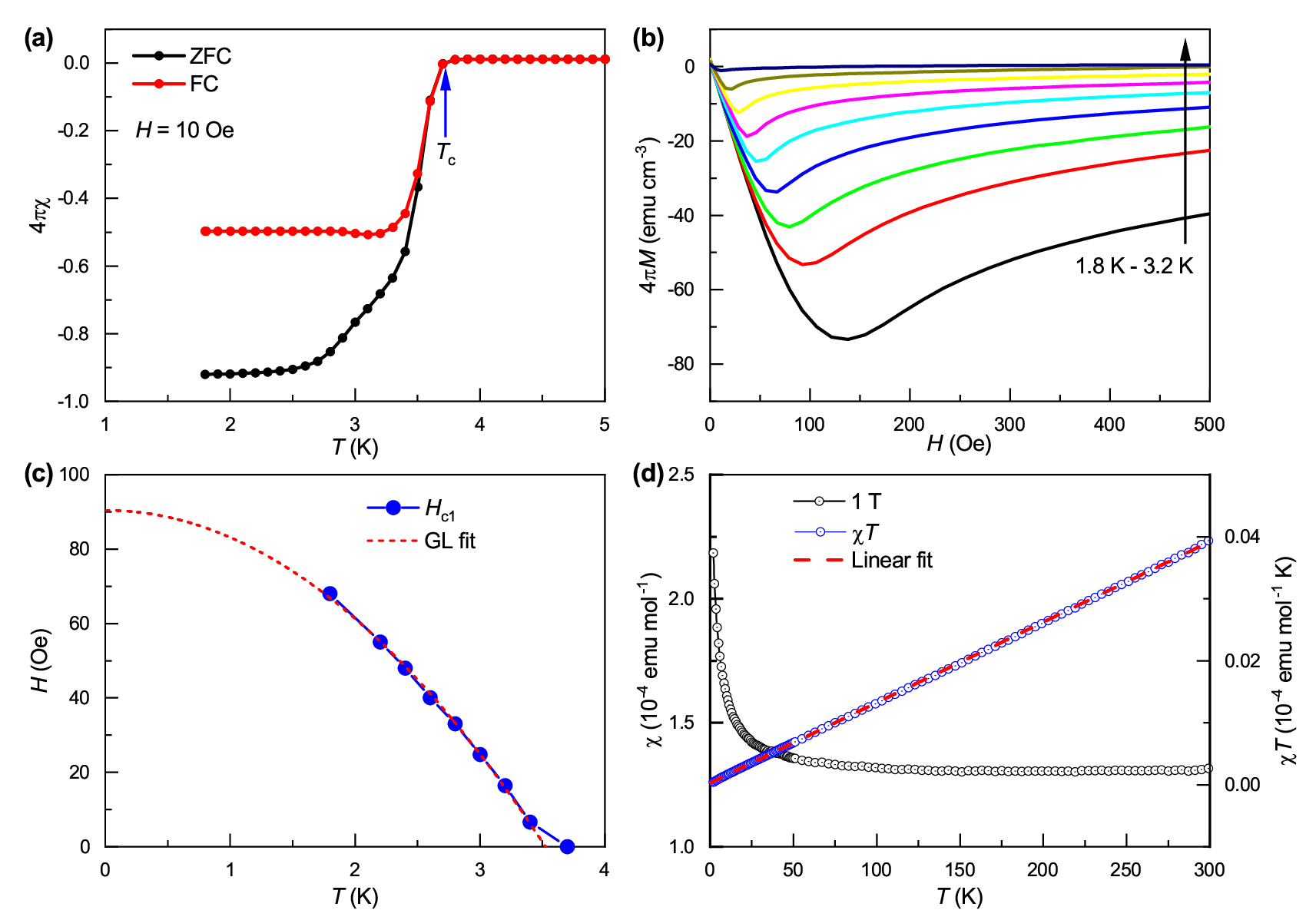}
\caption{(a) The low-temperature susceptibility for ThRu$_3$Si$_2$ in ZFC (black) and FC (red) modes under magnetic field of 10~Oe. (b) Field dependence of initial magnetization at various temperatures below $T_{\rm{c}}$. (c) Temperature dependence of the lower critical field $H_{\rm{c1}}$($T$). (d) Temperature dependence of magnetic susceptibility $\chi$ at 1~T. The right axis corresponds to $\chi$$T$, and the dashed red line is linear fit of $\chi$$T$.}
\label{Fig3}
\end{figure*}
The temperature dependence of  magnetic susceptibility was measured in zero-field cooling (ZFC) and field-cooling (FC) modes as shown in Fig.~\ref{Fig3}(a). The diamagnetism appears at about 3.7~K, which corresponds to the zero point in resistivity. The superconducting volume fraction approaches above $\sim$90$\%$ at 1.8 K in ZFC mode, indicating bulk SC here. The Meissner screening in FC mode decreases to only about half of that in ZFC mode. The weaker diamagnetic signal in FC mode indicated flux pinning in a type \uppercase\expandafter{\romannumeral2} superconductor. Fig.~\ref{Fig3}(c) shows the magnetization curves at different temperatures below SC. The lower critical fields $H_{c1}$ are determined from the fields where the $4\pi$$M$ starts to deviate from linearity. The temperature dependence of $H_{\rm{c1}}$ is fitted by the Ginzburg$\mbox{-}$Landau formula $H_{\rm{c1}}$ = $H_{\rm{c1}}$(0)[(1-($T/T_{\rm{c}}$)$^2$)]. The $H_{\rm{c1}}$(0), lower critical field at 0~K, is about 89.9(1) Oe.

The susceptibility under field of 1~T is nearly independent of temperature above 150~K, and the tail at low temperature is possibly induced by tiny paramagnetic impurities or defects, as present in Fig.\ref{Fig3}(d). To simplify, we consider the susceptibility to be described by $\chi$ = $\chi_0$ + $C/T$, where the term of $C/T$ is the contribution from the paramagnetic impurities or defects and the $\chi_0$ is a temperature$\mbox{-}$independent term. As shown in the right side of Fig.~\ref{Fig3}(d), the linear fitting of $\chi$$T$ versus $T$ yields $\chi_0$ = 1.30(4)$\times$10$^{-4}$ emu/mol and $C$ = 2.65(7)$\times$10$^{-4}$ emu/mol K. The constant $C$ corresponds to a small effective moment of 0.05 $\mu_{\rm{B}}$/f.u., demonstrating the extrinsic paramagnetism. The $\chi_0$ is considered to be contributed by several terms, $\chi_0$ = $\chi_{\rm{P}}$ + $\chi_{\rm{core}}$ + $\chi_{\rm{VV}}$ +$\chi_{\rm{L}}$, where $\chi_{\rm{P}}$ is the Pauli susceptibility and the other three terms corresponding to diamagnetic atomic$\mbox{-}$core susceptibility, the van Vleck paramagnetic susceptibility, and the Landau diamagnetic susceptibility, respectively. The $\chi_{\rm{core}}$ is estimated by the Pascal method with the value of 1.18$\times$10$^{-4}$ emu/mol~\cite{diam}, while the $\chi_{\rm{L}}$ is about -1/3$\chi_{\rm{P}}$. The $\chi_{\rm{VV}}$ usually takes place in rare$\mbox{-}$earth ions  with an even number of electrons in the unfilled 4$f$ shells, thus we ignore contribution of van Vleck magnetism here. Therefore, the $\chi_{\rm{P}}$ is estimated to be 3.72$\times$10$^{-4}$ emu/mol.

\begin{figure*}[htb!]
\centering
\includegraphics[width=1.5\columnwidth]{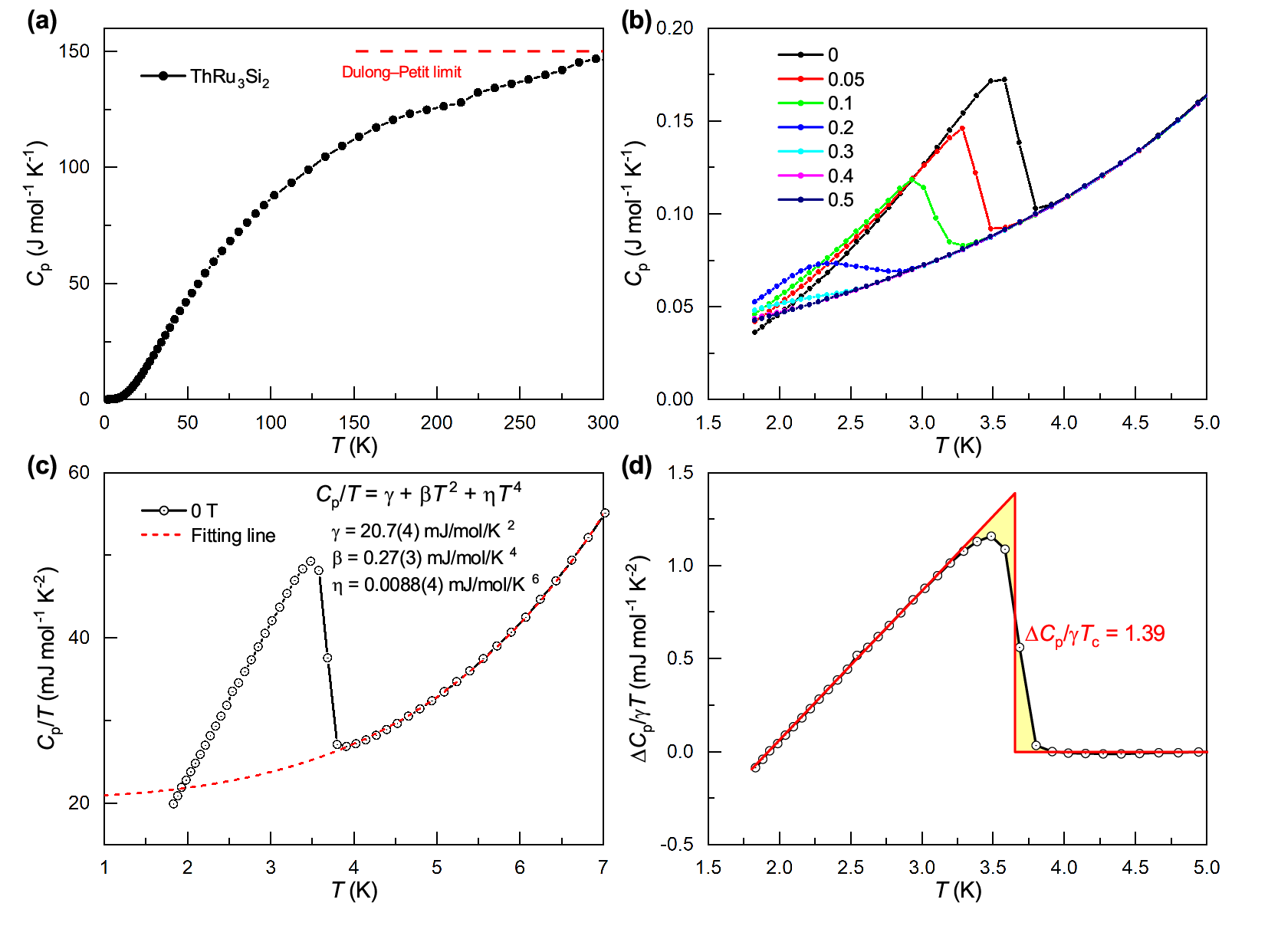}
\caption{(a) Temperature dependence of specific heat $C_{\rm{p}}$ for ThRu$_3$Si$_2$. (b) Low$\mbox{-}$temperature $C_{\rm{p}}$ under various fields. (c) Plot of $C_{\rm{p}}/T$ versus $T$ below 7~K. The dashed red line is the fit based on the formula $C_{\rm{p}}/T$ = $\gamma$ + $\beta$$T^2$ + $\eta$$T^4$. (d) The specific$\mbox{-}$heat contribution from SC by subtracting the electrons' and phonons' parts.}
\label{Fig4}
\end{figure*}

The temperature dependence of specific heat $C_p(T)$ is present in Fig.~\ref{Fig4}(a), which saturates at the Dulong$\mbox{-}$Petit limit of 3$NR$ = 150 J K$^{-1}$ mol$^{-1}$ at room temperature, where the $N$ is the number of atoms per formula unit and $R$ is the gas constant.
At low temperature, contribution from itinerant electrons is remarkable to specific heat. Fig.~\ref{Fig4}(b) shows the fitting of $C_{\rm{p}}$/$T$ below 7~K using the formula $C_{\rm{p}}/T$ = $\gamma$ + $\beta$$T^2$ + $\eta$$T^4$, where the $\gamma$ is Sommerfeld coefficient corresponding to specific heat
of electrons, and the two higher order terms is the contribution of phonons.~\cite{HC}  The fitiing gives $\gamma$ = 20.7(4) mJ K$^{-2}$ and Debye temperature $\theta_{\rm{D}}$ = 351(5)~K. The $\theta_{\rm{D}}$ is calculated by the formula $\theta_{\rm{D}}$ = ($\frac{12\pi^4}{5\beta}NR$)$^{1/3}$. A jump corresponding to the superconducting transition is observed at 3.8~K, which increases to maximum  at 3.6~K, corresponding a transition width of 0.2~K, which matches the sharp transition in resistivity. The transition shifts to lower temperature under external magnetic field, and is not observable above 1.8~K at 0.5~T. The specific heat jump at $T_{\rm{c}}$, $\Delta$$C$/$\gamma$$T_{\rm{c}}$ $\sim$ 1.39, further confirming bulk SC in ThRu$_3$Si$_2$. This value is  very close to 1.43, which is the theoretical value of BCS theory in the weak coupling limit. Based on the Debye temperature, the electron$\mbox{-}$phonon coupling constant $\lambda_{\rm{ep}}$ can be also estimated through McMillan equation~\cite{Mcm}:
\begin{equation}\label{1}
\lambda_{\rm{ep}} = \frac{1.04+\mu^{*}\rm{ln}(\frac{\theta_{\rm{D}}}{1.45\emph{T}_{\rm{c}}})}{(1-0.62\mu^{*})\rm{ln}(\frac{\theta_{\rm{D}}}{1.45\emph{T}_{\rm{c}}})-1.04}
\end{equation}
where the $\mu^{*}$ is the Coulomb pseudopotential parameter with a typical value of 0.13. The $\lambda_{\rm{ep}}$ value for ThRu$_3$Si$_2$ is 0.57, which is close to the values of 0.67 in LaRu$_3$Si$_2$($\theta_{\rm{D}}$ = 412~K and $T_{\rm{c}}$ = 7.8~K) and 0.5 in YRu$_3$Si$_2$~\cite{Y132}. Therefore, a moderate strength of electron$\mbox{-}$coupling is expected in the Ru based 132 system.

\begin{figure*}[t]
\centering
\includegraphics[width=1.5\columnwidth]{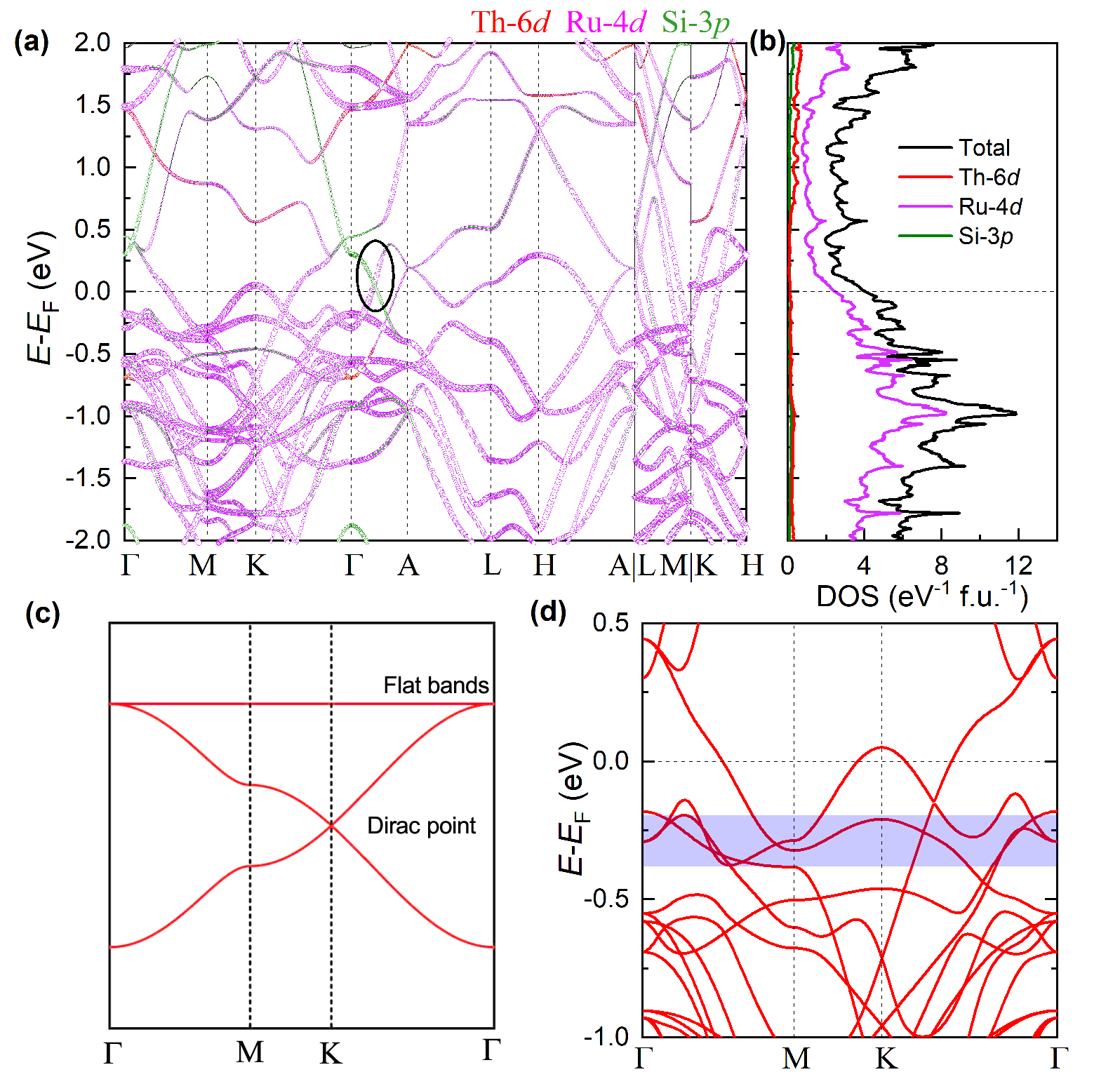}
\caption{(a)(b) DFT calculation of band structure and projected density of states without spin$\mbox{-}$orbital coupling. The black circle shows the possible Dirac point near the Fermi level.(c) Typical band structure of kagome lattice with Dirac cone at the K point and a flat band across the Brillouin zone. (c) Typical band structure of kagome lattice with Dirac cone at the K point and a flat band across the Brillouin zone. (d) The band structure near $E_{\rm{F}}$. The blue region highlights the kagome flat band.}
\label{Fig5}
\end{figure*}

To get more insights into the properties of ThRu$_3$Si$_2$, the density of electronic states and electronic band structure are calculated based on density$\mbox{-}$functional theory. Fig.~\ref{Fig5}(a) shows the projected band structure ThRu$_3$Si$_2$. There are two bands cross the $E_{\rm{F}}$, which is consistent with the two$\mbox{-}$bands fitting of the upper critical field $\mu_0H_{\rm{c2}}$($T$). The bands from $\Gamma$ to A shows prominent dispersion, indicating the two$\mbox{-}$dimensional feature is not remarkable in ThRu$_3$Si$_2$. Additionally, a possible Dirac point with the bands cross of Si 3$p$ and Ru 4$d$ electrons appears about 70~meV above $E_{\rm{F}}$, as shown in the circled area. The projected density of state (DOS) shown in Fig.~\ref{Fig5}(a) demonstrates the states around $E_{\rm{F}}$  are mainly contributed by the Ru 4$d$ electrons. The total DOS $N_{\rm{E}}$ at $E_{\rm{F}}$ is 4.06 eV$_{-1}$ f.u.$_{-1}$. This value is about one third of 11.78~eV$_{-1}$ f.u.$_{-1}$ inferred from Pauli susceptibility with the formula: ($\chi_{\rm{P}}$ = $\mu_{\rm{B}}^2$$N_{\rm{E}}$), and also less than half of 8.7 eV$_{-1}$ f.u.$_{-1}$ obtained from Sommerfeld coefficient through the formula $\gamma$ = $\frac{1}{3}\pi^2$$k_{\rm{B}}^2$$N_{\rm{E}}$, indicating the contribution  of quasiparticle interactions in itinerant electrons. Wilson ratio is usually used to estimate the electron correlation effect, which is calculated with the equation: $R_{\rm{W}}$ = $\frac{1}{3}$$\frac{\pi k_{\rm{B}}}{\mu_{\rm{B}}}^2$$\frac{\chi_{\rm{P}}}{\gamma}$. The obtained $R_{\rm{W}}$ is 1.31, slight larger than 1, indicating the existence of correlation of quasi$\mbox{-}$particles.

Fig.~\ref{Fig5}(d) shows the band structure around $E_{\rm{F}}$. A Dirac point at the K point with linear dispersion is observed at 700~meV below the $E_{\rm{F}}$. And the van Hove point is also found away from  $E_{\rm{F}}$ (300 meV) at M point. Besides, the flat bands, which are common in  two$\mbox{-}$dimensional kagome lattice from $\Gamma$ to M point as shown in Fig.~\ref{Fig5}(c), present severe dispersion.  Compared with the results of LaRu$_3$Si$_2$, despite of the slight difference in the details of the energy bands, the $E_{\rm{F}}$ is basically lifted up about 400~meV in ThRu$_3$Si$_2$, which can be considered as electron doping. The flat bands are farther away from the $E_{\rm{F}}$ in ThRu$_3$Si$_2$ (300~meV above $E_{\rm{F}}$) than that in LaRu$_3$Si$_2$ (100~meV below $E_{\rm{F}}$).

\section{\label{sec:level4} Conclusions}
\begin{table}[h]
\footnotesize
\caption{Physical property parameters of LaRu$_3$Si$_2$ and ThRu$_3$Si$_2$.}
\label{property}       
\tabcolsep 3pt
\renewcommand\arraystretch{1.4}
\begin{tabular}{lcc}
\hline\noalign{\smallskip}
Parameters & LaRu$_3$Si$_2$~\cite{2011La132} & ThRu$_3$Si$_2$\\
\noalign{\smallskip}\hline\noalign{\smallskip}
$a$~($\rm{{\AA}}$) & 5.68 & 5.614 \\
$c$~($\rm{{\AA}}$) & 7.13 & 7.191 \\
$T_{\mathrm{c}}$~(K)  &7.8 & 3.8\\
RRR  &5 & 7\\
$\chi_{\rm{P}}$~($10^{-4}$~emu mol$^{-1}$ Oe$^{-1}$) &14.4 &3.72\\
$\gamma$~(mJ K$^{-2}$ mol$^{-1}$) & 36.8 & 20.7\\
$\theta_{\rm{D}}$~(K) & 412 & 351\\
$\lambda_{\rm{ep}}$ &0.67&0.57\\
$\Delta$$C$/$\gamma$$T_{\rm{c}}$ &1.36&1.39\\
$R_{\rm{W}}$ &2.88&1.31\\
$N_{\rm{E}}$~(eV$^{-1}$ f.u.$^{-1}$) &$\sim$7& 4.06\\
\noalign{\smallskip}\hline\noalign{\smallskip}
\end{tabular}
\end{table}
For comparison, we list the physical parameters of LaRu$_3$Si$_2$ and ThRu$_3$Si$_2$ in Table~\uppercase\expandafter{\romannumeral1}. There are no much difference in the cell parameters $a$ and $c$, which can be ascribed to the similar ion radius of La$^{3+}$ and Th$^{4+}$. However, the critical temperature of superconductivity is doubled in LaRu$_3$Si$_2$, which should mainly originate from the significant contrast in density of states near the Fermi level, as revealed in the DFT calculations. The Fermi level is further away from the flat bands. Additionally, the stronger electron$\mbox{-}$phonon coupling, which is proportional to the Debye temperature, also contributes higher $T_{\rm{c}}$ in LaRu$_3$Si$_2$. Similar to LaRu$_3$Si$_2$~\cite{2011La132,2021La132,NQRLa132,DopingLa132}, ThRu$_3$Si$_2$ is also considered to be moderately coupled BCS superconductor according to the values of $\lambda_{\rm{ep}}$. The specific heat jump at $T_{\rm{c}}$ in the two compounds are close to the theoretical value 1.43. In ThRu$_3$Si$_2$,  the upper critical field $\mu_0H_{\rm{c2}}$($T$) can be well fitted by a two band model, which is also consistent with the multi$\mbox{-}$band feature shown in the DFT calculation.

Interestingly, the total DOS at $E_{\rm{F}}$ estimated from the specific$\mbox{-}$heat coefficient $\gamma$ and Pauli paramagnetism $\chi_{\rm{P}}$, are much larger than that obtained by DFT calculation, suggesting a electron$\mbox{-}$mass renormalization due to correlation effect.  This renormalization is more prominent in LaRu$_3$Si$_2$. On the other hand, the Wilson ratio $R_W$, which is equal to 1 for  noninteracting free electron gas system, are 2.88 and 1.31 for LaRu$_3$Si$_2$ and ThRu$_3$Si$_2$ respectively, also suggesting the existence of electron correlations and it is stronger in LaRu$_3$Si$_2$.  The correlation effect is considered to arise from the flat bands. We note that the Fermi level is shifted from 100~meV below flat bands in LaRu$_3$Si$_2$ to 300~meV above it in ThRu$_3$Si$_2$. The effect of flat bands is weakened as it away from the Fermi level, accompanied by the decrease of density of states and suppression of electron correlations.

In summary, the bulk superconductivity of kagome metal ThRu$_3$Si$_2$ with $T_{\rm{c}}$ = 3.8~K are investigated through resistivity, magnetism and specific heat measurements. The temperature dependence of upper critical field suggests two superconducting gaps in ThRu$_3$Si$_2$ with $H_{\rm{c2}}$(0) = 1.45~T, which is well below the Pauli limit. The specific heat jump and calculated electron$\mbox{-}$phonon coupling indicate a moderately electron$\mbox{-}$phonon coupled superconductor. On the other hand, affected by the flat bands, the electron correlation in ThRu$_3$Si$_2$ is relatively weakened compared with LaRu$_3$Si$_2$. Since the Fermi level in ThRu$_3$Si$_2$ shifts up about 400~meV arising from the introduction of extra electrons by substituting La with Th, the flat bands are further away from the Fermi level, which explains the reduction of DOS and electron correlations. Therefore, it is expectable that higher  $T_{\rm{c}}$  and stronger electron correlations may appear in Th slightly doped LaRu$_3$Si$_2$ with flat bands right on the Fermi level.

{$^{\ast}$ Yi Liu and Jing Li contribute equally to this work.}\\
\begin{acknowledgments}
This work is supported by the National Natural Science Foundation of China (Grant No. 12050003,No. 12004337 and No. 12274369) and Zhejiang Provincial Natural Science Foundation of China (Grants No.LQ21A040011).
\end{acknowledgments}

\bibliographystyle{iopart-num}
\providecommand{\newblock}{}


\end{document}